\documentclass[10pt]{article}
\def\ba{\begin{eqnarray}}
\def\ea{\end{eqnarray}}

\usepackage{graphicx}
\usepackage{amssymb}

\begin{document}
\title{\Huge\bf Geometrical pinning of magnetic vortices induced by
a deficit angle on a surface: anisotropic spins on a conic space background}
\author{{W.A. Moura-Melo}$^{1,\,}$\thanks{E-mail: winder@ufv.br} , {A.R. Pereira}$^{1,2,\,}$\thanks{E-Mail: apereira@ufv.br} ,
{L.A.S. M\'ol}$^{3\,}$\thanks{E-mail: lucasmol@fisica.ufmg.br} , and A.S.T.
Pires$^{3}$ \thanks{E-mail: antpires@fisica.ufmg.br}\\ \\
\small
$^{1}$ \it Departamento de F\'{\i}sica, Universidade Federal
de Vi\c{c}osa\\ \small \it 36570-000, Vi\c{c}osa, Minas
Gerais, Brazil.\\
\small
$^{2}$ \it Physics Department, University of Bolonga\\ \small \it Via Irnerio 46, I-41126, Bolonga, Italy.\\
\small $^{3}$ \it Departamento de F\'{\i}sica, ICEX, Universidade Federal de
Minas Gerais\\ \small \it  Caixa Postal 702, 30123-970, Belo Horizonte, Minas
Gerais, Brazil}

 \date{}
\maketitle
\begin{abstract}
We study magnetic vortex-like excitations lying on a conic space
background. Two types of them are obtained. Their energies appear to be linearly dependent
on the conical aperture parameter, besides of being logarithmically
divergent with the sample size. In addition, we realize a
geometrical-like pinning of the vortex, say, it is energetically
favorable for it to nucleate around the conical apex. We also study the problem of two vortices
on the cone and obtain an interesting effect on such a
geometry: excitations of the same charge, then repealing each other,
may nucleate around the apex for suitable cone apertures. We also
pay attention to the problem of the vortex pair and how its
dissociation temperature depends upon conical geometry.
\end{abstract}


\section{Introduction and Motivation}
\indent Non-linear excitations play important role for
understanding several properties of physical systems. For
instance, a number of phase transitions observed in Cosmology,
High Energy, and Condensed Matter Physics are accompanied by such
kind of excitations. Particularly, in low dimensional magnetic
materials, such excitations develop important roles. This is the
case for two-dimensional spin models with planar symmetry, where a
topological phase transition takes place due to vortex-antivortex
dissociation \cite{B,KT}. Actually, such a transition
(Berezinskii-Kosterlitz-Thouless transition) is expected to occur
in a variety of systems  such as films of superfluid helium,
superconducting materials, Josephson-junctions as well as certain
magnetic, gaseous and liquid-crystal systems. Whether a similar
scenario would take place in other surface backgrounds (say,
curved support) remains still an open question. Of course, a first
step in this direction is to know how vortices behave in
non-Euclidian spaces. Concerning this, we should emphasize that magnetic materials structured in
curved
geometry have attracted a great deal of efforts in the last years.
Such an interest partially lies on the fact that magnetic samples
can be currently manipulated at very small scales (around some
dozens of nanometers or even smaller) and/or several
shapes\cite{spin-shapes}. The knowledge of their magnetic
properties is important for many potential applications, like
devices for data recording and logic gates\cite{spin-mag-aplic}.
Therefore, a number of works has been devoted to study curved
magnetic surfaces like cylinders, spheres, cones, and so forth.
Several results that emerge from the interplay between magnetic
and geometric effects have been obtained. For example, on the
surface of a magnetic cylinder, solitons appear to be
sine-Gordon-like excitations\cite{spin-cylinder}, which tend to
deform the cylindrical surface in order to relieve the geometrical
frustration brought about by nonconstant curvature, anisotropies,
Zeeman effect, etc\cite{spin-cylinder2}. Now, on a conical support
their energy gets lower as long as cone is narrowed, indicating
that such surfaces could be thought as pinning defects for
solitons\cite{spin-cone} (see, however\cite{SD-cone}). In addition,
transition from {\em flower} to {\em vortex-like} magnetization
has been experimentally observed in ferromagnetic nanosized cones
\cite{Ross-nanocone,Rossetal}. Therefore, investigation of the basic properties of such excitations in conical support is important not only from a fundamental point of view, it is also appealing from applied physical grounds. Moreover, it should be emphasized
that even `more exotic' manifolds, like those with negative
curvature (hyperbolic spaces) have been quite recently studied in
connection
with Condensed Matter and Statistical Physics \cite{shima}.\\

Here, our attention is focused on vortex-like excitations lying on
a circular conical surface covered by an anisotropic classical
Heisenberg magnet. Our initial motivation for considering such a
problem was to analyze how lattice defects (here a deficit angle)
affect excitations whose solution is angle-like dependent. Such
spatial defect is particularly interesting because it makes the
space locally Euclidian while it presents nontrivial global
curvature effects (due to the boundary conditions which identify
opposite points of the wedge cut out of space, as shown in Figure
\ref{cone}). Indeed, conic space can be described by a spatial
metric with a Riemann-Christoffel curvature tensor which vanishes
everywhere, except in one point, where there is a conical
singularity characterized by a deficit angle\cite{lowerdgravity}.
In a planar magnetic material, such a (area) deficit could be artificially
built without cutting out any part of the space, rather it could be obtained substituting the magnetic atoms by nonmagnetic sites in the large
region covered by the deficit angle. In addition, the spins placed at
opposite sites of the wedge should be identified. While this complex
construction should be a very difficult task nowadays (mainly
because of the nontrivial boundary conditions), this physical
system could be exactly reproduced in a simpler manner: a conical
support with appropriate magnetic coating. We shall see in what
follows that, by virtue of the deficit angle, vortex excitations
present lower energy than their counterpart
on a standard flat plane. Actually, energy diverges
logarithmically as sample conic area blows up, but decreases
linearly as cone aperture angle, $2\alpha$, is lowered. Here, we distinguish between two types of
vortices: {\em in-cone} and {\em out-of-cone}. The main difference between them lies in
the fact that the latter has a core where spins present components out of the surface. We also
relate the appearance of each type with conical aperture and location on the cone. Such issues will
be the subject of Section 2. Results from this section open the possibility of
a geometrical controlling of the vortex-like excitations energy on
actual finite-size samples and of pinning them around conically
shaped structures. Indeed, we shall realize that the magnetic
vortex tends to nucleate around the conical apex [Section 3].\\

We also address the problem of two vortices on the conical magnet.
This shall be done in Section 4. Similarly to the usual flat plane
case, a vortex pair on the conical surface presents finite energy,
whose value appears to increase logarithmically with the
separation of the vortices while decreasing linearly as the cone
is narrowed. An interesting result emerging from this section is the possibility of vortices with
the same charge, then mutually repealing, of nucleating around the apex. This is carried out for
two distinct arrangements of the excitations and for both cases we determine the suitable conical
apertures supporting the nucleation.
Following, we analyze possible consequences of the curvature on
the problem of depairing of vortices, namely, how critical temperature is
sensitive to geometrical parameters. The conical surface is,
perhaps, the first step in this direction since it is locally
Euclidian and, therefore, its geometry keeps some similarities
with the standard and well studied flat plane. Finally, we close
this work by pointing out our conclusions and prospects for
future investigation.

\section{The model and vortex-like solutions}
Let us consider an anisotropic Heisenberg-like model for
nearest-neighbors interacting spins on a two-dimensional lattice,
like below: \ba H_{1}=-\frac{J}{2}\,\sum_{<i,j>} {\cal
H}_{i\,,j}=-\frac{J}{2}\,\sum_{<i,j>}
(S^x_iS^x_j+S^y_iS^y_j+\lambda \, S^z_iS^z_j)\,, \label{Hlattice}
\ea where $J>0$ is the ferromagnetic exchange coupling between
neighbor spins and ${\vec{S}}_i=(S^x_i,S^y_i,S^z_i)$ is the spin
operator at site $i$. Parameter $\lambda$ answers for the
anisotropy between spin couplings: for $\lambda>1$ spins tend to
align along the $z$-axis (easy-axis regime); for $\lambda=1$ we
have the isotropic case; while for $0<\lambda<1$ one gets the
easy-plane regime. Finally, $\lambda=0$ leads to the so-called
$XY$ model.

Whenever dealing with a general surface, we could also take into
account another model which incorporates possible anisotropy due
to the alignment of spins along the normal of this surface, like
follows:
\begin{equation}
H_{2}=-\frac{J}{2}\,\sum_{<i,j>}
{\cal{H}}_{i\,,j}+\frac{J\delta}{2}\,\sum_i \,
(\vec{S}_i\cdot\hat{n}_i)^2, \label{Hlattice2}
\end{equation}
where $\hat{n}_i$ is a unity vector normal to the surface at each
site, $i$. Thus, for $\delta >0$ the spins tend to be locally
parallel to the surface (we shall take $\delta\ge0$, hereafter). This may be particularly
important for
taking into account, for instance, magnetostatic energy associated
with topological excitations lying on a finite and non-planar
background. In fact, $\delta$-term works like a single-ion
anisotropy parameter in a general surface, so that as long as we
are dealing with a flat one, say, $xy$-plane, then such a term
recover the usual single-ion anisotropy, $\delta \,(S^z_i)^2$.
Results concerning such an issue and vortex stability in (flat)
square, hexagonal and triangular lattices are available in the
works of Refs. \cite{Wysinprb54,Wysinpla1998}.\\

In the continuum approach of spatial and spin variables,
which is valid at sufficiently low temperature and long-wavelength limit,
Hamiltonian above may be expanded to yields:
\begin{equation}
 H=E_0-\frac{J}{2}\int\int_\Omega {\cal H}_{\rm
cont}d\Omega=E_0+\frac{J}{2}\int\int_\Omega \left[ ({\bf
D}\,\vec{S})^2 + \frac{\delta}{a^2} (\vec{S}\cdot\hat{n})^2\right]
\,d\Omega\,, \label{Hcont}
\end{equation}
where $E_0$ is the ground state energy, $a$ is the lattice spacing parameter, and we have used, for
simplicity $\lambda=1$. In turn, the classical spin state is taken
to be $\vec{S}=(\sin\theta\cos\Phi, \,\sin\theta\sin\Phi,\,\cos
\theta)$, with $\theta=\theta(\vec{x},t)$ and
$\Phi=\Phi(\vec{x},t)$. Here, $\hat{n}$ is an unity vector field
normal to the surface at each point. Therefore, the
$S^2$-parametrization of $\vec{S}$ is such that, at each point of
the physical surface, $\theta$ is the angle between $\vec{S}$ and
$\hat{n}$, $\vec{S}\cdot\hat{n}=\cos(\theta)\equiv\,m$, while
$\Phi$  is the local polar angle. In turn, the integral is taken
over an arbitrary surface, $\Omega$, while ${\bf D}=\sum_i
\hat{e}^i D_i$ represents the covariant derivative on such a
space. Recalling that $\vec{S}^2=1$, the term $\int\int_\Omega
({\bf D}\,\vec{S})^2 d \Omega$ in Hamiltonian above is the
nonlinear $\sigma$ model in a curved space background. Therefore,
Hamiltonian (\ref{Hcont}) describes a kind of anisotropic
nonlinear $\sigma$ model on a curved surface.

\begin{figure}[!h]
\centering \hskip 1cm 
\fbox{\includegraphics[width=15cm,height=7cm]{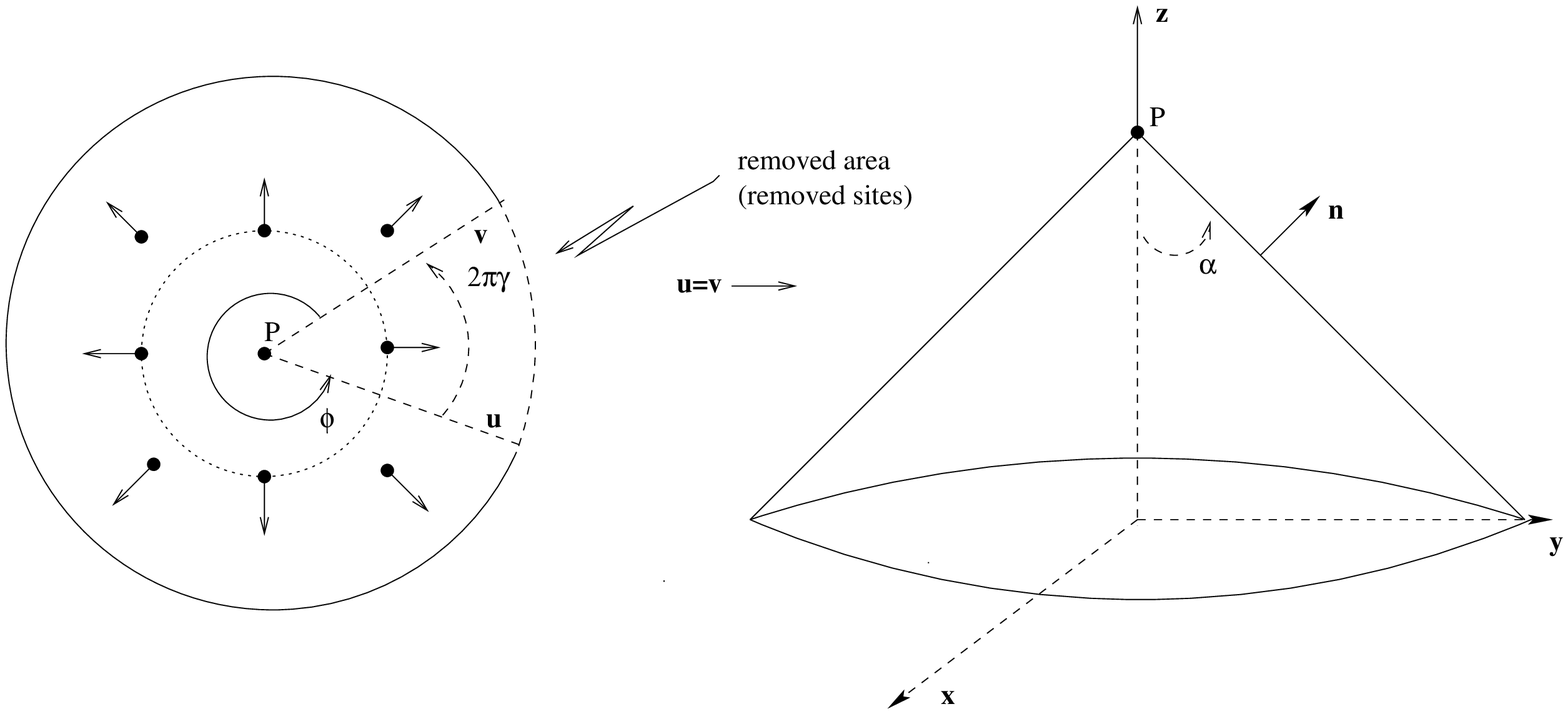}} \vskip
-0.0cm \caption{{\protect\small Here, it is shown cut and removal of a
wedge-type area equal to $2\pi\gamma\,r^2$ from the original disc
area, $2\pi\,r^2$. Removing magnetic sites, now as $\phi$ goes
from $0$ to $\pm 2\pi(1-\gamma)=\pm2\pi\beta$ the identification of lines {\bf u} and {\bf v}
geometrically forces the spins to take a complete round. Note that $\beta=\sin(\alpha)\,\in (0,1)$, for usual cones. [For improving visualization, arrows
representing spins are not depicted on the cone].}} \label{cone}
\end{figure}

As it is well-known, the main ingredient to obtain a vortex solution is a field
equation that reduces in some limit to a Laplacian yielding the
energy density $(\vec{\nabla}\Phi)^{2}$, where $\Phi$ is the
phase. Such a limit is implicit in our system as one can easily
see below. First, due to the local Euclidian nature of the conical
surface, it is convenient to use local flat coordinates
\cite{spin-cone,lowerdgravity} defined as
$\vec{\rho}=(\rho,\tau)=(r^{\beta}/\beta,\,\beta \phi)$, where
$\vec{r}=(r,\phi)$ are assumed to be the usual polar coordinates. Thus we
can rewrite Hamiltonian (\ref{Hcont}) as follows:
\begin{equation}
H=\frac{J}{2}\int_{0}^{\infty}\int_{0}^{2\pi
\beta}\left[\frac{(\vec{\nabla}_{\vec{\rho}}\,
m)^{2}}{1-m^{2}}+(1-m^{2})(\vec{\nabla}_{\vec{\rho}}\, \Phi)^{2}+
\frac{\delta}{a^{2/\beta}_\rho} \beta^{-2}\, \rho^{\frac{2-2\beta}{\beta}} \, m^{2}
\right] \rho d\rho d\tau ,\label{HCont2}
\end{equation}
where $\vec{\nabla}_{\vec{\rho}}$ is the gradient written in terms
of coordinates $(\rho,\tau)$ and $a_\rho=a^\beta /\beta$ is
the analog of the actual lattice spacing parameter $a$, written in $(\rho,\tau)$ coordinates. Note
that a spin
lying on this geometry experiences an effective anisotropy
$\Delta=\delta \,\beta^{-2} \,\rho^{\frac{2-2\beta}{\beta}}$ that depends
on its position (site) $\rho$ on the surface as well as the cone
aperture, measured by the parameter $\beta=\sin(\alpha)$, $\alpha$ being the cone half-angle (see Figure \ref{cone}).\\
The second step for getting the vortex solutions
is to obtain the equations of motion on this support. Their static counterparts can be found from eq. (\ref{HCont2}), and read like follows:
\begin{eqnarray}
& & \nabla_{\vec{\rho}}^{2}\, m+
\frac{m(\vec{\nabla}_{\vec{\rho}}\, m)^{2}}{1-m^{2}}+m(1-m^{2})\left[(\vec{
\nabla}_{\vec{\rho}}\, \Phi)^{2}-\frac{\Delta}{a_\rho^{2/\beta}}\right]=0
,\label{EMotion1}\\
& & \nabla_{\vec{\rho}}^{2}\Phi -\frac{2m(\vec{\nabla}_{\vec{\rho}}
m)\cdot (\vec{\nabla}_{\vec{\rho}} \Phi)}{1-m^{2}}=0
,\label{EMotion2}
\end{eqnarray}
whose solutions must obey the auxiliary conditions $m(0)=m(2 \pi
\beta)$ and $\Phi(0)=\Phi(2\pi \beta)$ due to the global curvature
effects. This global effect, i.e., the identification of
opposite points of the wedge is fundamental for the existence of
vortices in this geometry. With these conditions, one can see that Hamiltonian (\ref{Hcont}) and
the equations of motion (\ref{EMotion1}) and (\ref{EMotion2}) now have all necessary
ingredients to produce a vortex solution. A static vortex (for a
while assumed to be centralized at the conical apex) is the pair
of functions
$(m,\Phi)=(\,f(\rho\,\sqrt{\Delta}),\,
Q g(\vec{\rho}))$ that satisfy Eqs. (\ref{EMotion1}) and
(\ref{EMotion2}). Here $Q=\pm1,\pm2,\ldots$ is the topological charge, positive for
vortices and negative for antivortices. Away from the vortex
center the field $g(\vec{\rho})$ is proportional to the angle
$\tau$ and has the form $g=\tau$ (or $\Phi=Q\tau$). The field $m(\rho)$ must regularize the vortex
at the core, but it has a strong dependence on the anisotropy, $\delta$, and on the cone aperture,
$\beta$, as well. In the usual flat
space, this field can be obtained only numerically and the
analysis leads to two different types of vortices known as
in-plane and out-of-plane vortices, depending on whether the static
vortex has zero or nonzero out-of-plane spin components \cite{Wysinpla1998},
respectively. In this case, the out-of-plane
region is restricted to a small core region (of size
$l_{0}=a/\sqrt{\delta}$) at the vortex center. Here, in contrast
to the standard flat space case, spins placed at sites with
different values of $\rho$ experience different anisotropies. We should stress that such a feature
is not characteristic of the material itself, once $\delta$ was assumed to be the same at all
sites, rather it is related to the underlying geometry of the magnet.
Indeed, for any $\delta$ and $\beta\neq 1$, the effective
anisotropy vanishes in the limit $\rho\rightarrow 0$, and
therefore, the spin system behaves like an (almost) isotropic
magnetic material around the conical apex. Hence, independently of
the values of $\delta$, ``out-of-cone" spin components could also
develop around the tip of the cone. On the other hand, as $\rho$
increases, the effective anisotropy grows considerably so that
the spins are expected to be confined at the conic surface. Thus, out-of-cone fluctuations are
not expected to be observed far away from the tip.\\
 
Nevertheless, based on the above discussion and
on the results for soliton configurations on the conical surface
(see Ref.\cite{spin-cone}), we also expect two types of vortex
solutions, which will be referred to as `in-cone` and `out-of-cone`
vortices. Of course, it would be energetically favorable for
out-of-cone vortices to nucleate preferentially around the cone
apex, while, in contrast, in-cone vortices would prefer to
nucleate away from this region. For the
particular case of a vortex centered at the cone apex, results of
Ref.\cite{spin-cone}, concerning the problem of an isotropic
magnetic system on a cone, suggest that $m(0)=\pm 1$, while the
term $\Delta=\delta \,\beta^{-2}\, \rho^{\frac{2-2\beta}{\beta}}$ should
imply $m(\rho)\rightarrow 0$ for $\rho \rightarrow \infty $.
Really, the term $\int (\Delta m^{2})\rho d\rho$ in Hamiltonian
(\ref{Hcont}) diverges if $ m\neq 0$ at $\rho \rightarrow
\infty$. Consequently, the configuration of an out-of-cone vortex,
which may nucleate only around the cone tip, can be summarized as follows: $\Phi=Q\tau$, $m(0)=p$,
$m(\rho)\rightarrow
0$ for $\rho\rightarrow \infty$, where $p=\pm 1$ is the so-called
polarity of the vortex. The size of the out-of-cone region can be
estimated, in analogy to $l_0=a/\sqrt{\delta}$, if we take $l_{\rho}(\beta,\delta)=e^\beta\,a_\rho \,\beta^{\beta/(1-\beta)}/\sqrt{\delta}$. For
a fixed $\delta$, $l_\rho$ grows
considerably whenever $\beta$ is decreased. Then, out-of-cone magnetic interactions play
important roles as the conical apex becomes sharper. However, it is reasonable to assume that such interactions can be
neglected when the surface is nearly planar and/or
the values of $\delta$ are sufficiently large. In contrast to
``out-of-cone" vortices, the ``in-cone" vortices, characterized by
a core with spins confined at the conical surface, have more
chance of nucleating away from the cone tip, i.e., their centers
may not be found at $\rho =0$ (unless, of course, $\delta >>0$
and/or $\beta \approx 1$). Besides, it is also interesting to note
that such excitations can have relatively strong out-of-cone spin
fluctuations around the conical apex.\\

\indent Here, we shall firstly specialize to cylindrically
symmetric solutions on the surface of a circular cone, say,  $m=m(\rho)$ and $\Phi=\Phi(\tau)$ with
 unity topological charge, $Q=\pm1$. In this case, asymptotic
solutions for out-of-cone vortices can be obtained. The region
defined by $\rho<l_{\rho}(\beta,\delta)$ is nearly
isotropic and the $\Delta$-term can be neglected. Hence,
based on results of Ref.\cite{spin-cone} one can easily find the
out-of-cone vortex configuration for $\rho << l_{\rho}(\beta,\delta)$
\begin{equation}
m_{oc}\cong -p
\left(1-\frac{\rho^{2}}{l_{\rho}^{2}}\right),  \qquad
\Phi_{oc}=\tau. \label{outcone1}
\end{equation}
For the outer region $\rho >> l_{\rho}(\beta,\delta)$, one has
$m_{oc}=0$ and $\Phi_{oc}=\tau$. The exchange energy of such a solution may be estimated, using eq. (\ref{HCont2}), to be:
\ba
E_{\rm oc}\approxeq
\pi\,J\,\beta\,\ln\left(\frac{L}{l_c}\right)+2\pi\beta{J}\left[\frac34
+2\ln(2) +\frac{\beta\delta}{(\beta+1)(2\beta+1)}\frac{l^2_c}{a^2}\right]\,,\label{Eoc}
\ea
where $L$ and $l_c$ ($l_c=(\beta\,l_\rho)^{1/\beta}=(e\,a\,\beta^{1/(1-\beta)}/\delta^{1/2\beta}$)
represent the cone size and the out-of-cone core region size, respectively (in units of the actual
lattice
spacing,$a$).  Clearly, in the limit of an infinite cone, the first term blows up; the second one,
taking into account the contribution from the core, is finite. Notice also that, the cone geometry
plays an
important role in the energy behavior, so that it linearly decreases with conical aperture,
vanishing in the limit $\beta\to0$, even if $L\to\infty$. For arbitrary $Q$ energy above
is proportional to $Q^2$.\\

On the other hand, whenever $l_{c}(\beta,\delta)\rightarrow 0$ (strong anisotropy
and $\beta$ large enough), in-cone vortex solutions given by
($m_{ic}=0, \Phi_{ic}=\tau$) must become stable, even at the tip. In this case, the in-cone vortex
energy reads:
\ba
E_{\rm ic}=\pi\,JQ^2\beta\ln(L/a_0)\,,\label{Eic}
\ea
where $a_0$ is a cutoff introduced to prevent spurious divergences associated to the continuum
approach and also to provide a
correct value for the exchange energy. For example, for an in-plane vortex on a planar square
lattice we have that $a_0\approx 0.24a$\cite{Wysinprb54}. Similarly to the out-of-cone solution,
also here energy linearly decreases with $\beta$-parameter, while diverging like $\ln(L)$. A
detailed study of how the
field $m$ depends on the vortex position and how vortex stability
depends on the anisotropy and curvature is more involved and we
plan to do this in the near future. In addition, note that since the energies (\ref{Eoc}) and (\ref{Eic}) differ by a finite term, then results obtained for one kind of vortex could be, at some extent, applied to the another.\\

The main conclusion is the following: whenever lying on the
surface of a cone, a vortex-like excitation presents lower
exchange energy than its flat plane counterpart. Moreover, as cone
is narrowed, such an energy linearly decrease. We may
understand this in a simple geometric way, like follows: a cone with a narrower
opening angle corresponds to a smaller magnet, which therefore has
less magnetic energy. \\

It should be noticed that Belavin-Polyakov-like solitons present a
similar characteristic whenever defined on a conical
support\cite{spin-cone}, say, their energy also linearly decreases
with $\beta$ parameter (see, however, Ref.\cite{SD-cone}, where is
considered another solution, on a truncated cone, whose energy does not depend on the
conical aperture angle). Furthermore, we shall see, in the next
section, that vortex energy is lower whenever it nucleate around
conical apex, a fact that could be interpreted as a geometrical
pinning of magnetic vortex.

\section{The geometrical pinning of magnetic vortices on a conical support}

\begin{figure}[ht]
\centering \hskip 0cm 
\fbox{\includegraphics[width=15cm,height=8cm]{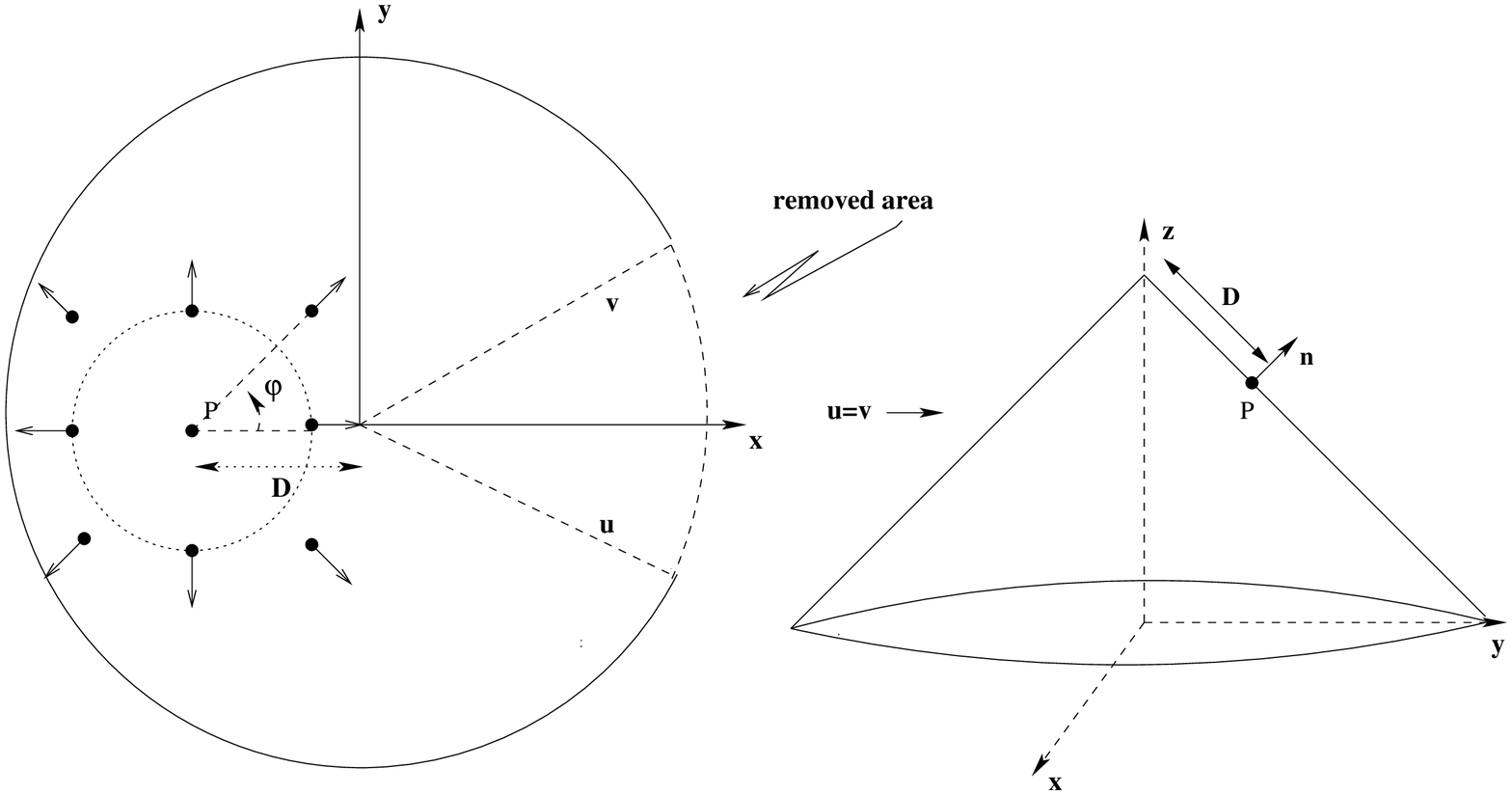}} \vskip
-0.0cm \caption{{\protect\small Similar to the preceding Figure.
Here, however, vortex is centered at a point $P$ which is apart a
distance $D$ from the conical apex. Then, note that the energy may be split in two terms, like eq. (\ref{EnergyD}), so that the term associated to the disc of radius $D$ experiences no angle deficit. [On the cone, arrows
representing spins are not depicted to improve visualization].}} \label{cone2}
\end{figure}

\begin{figure}[!h]
\centering \hskip 0cm
\fbox{\includegraphics[width=14cm,height=7cm]{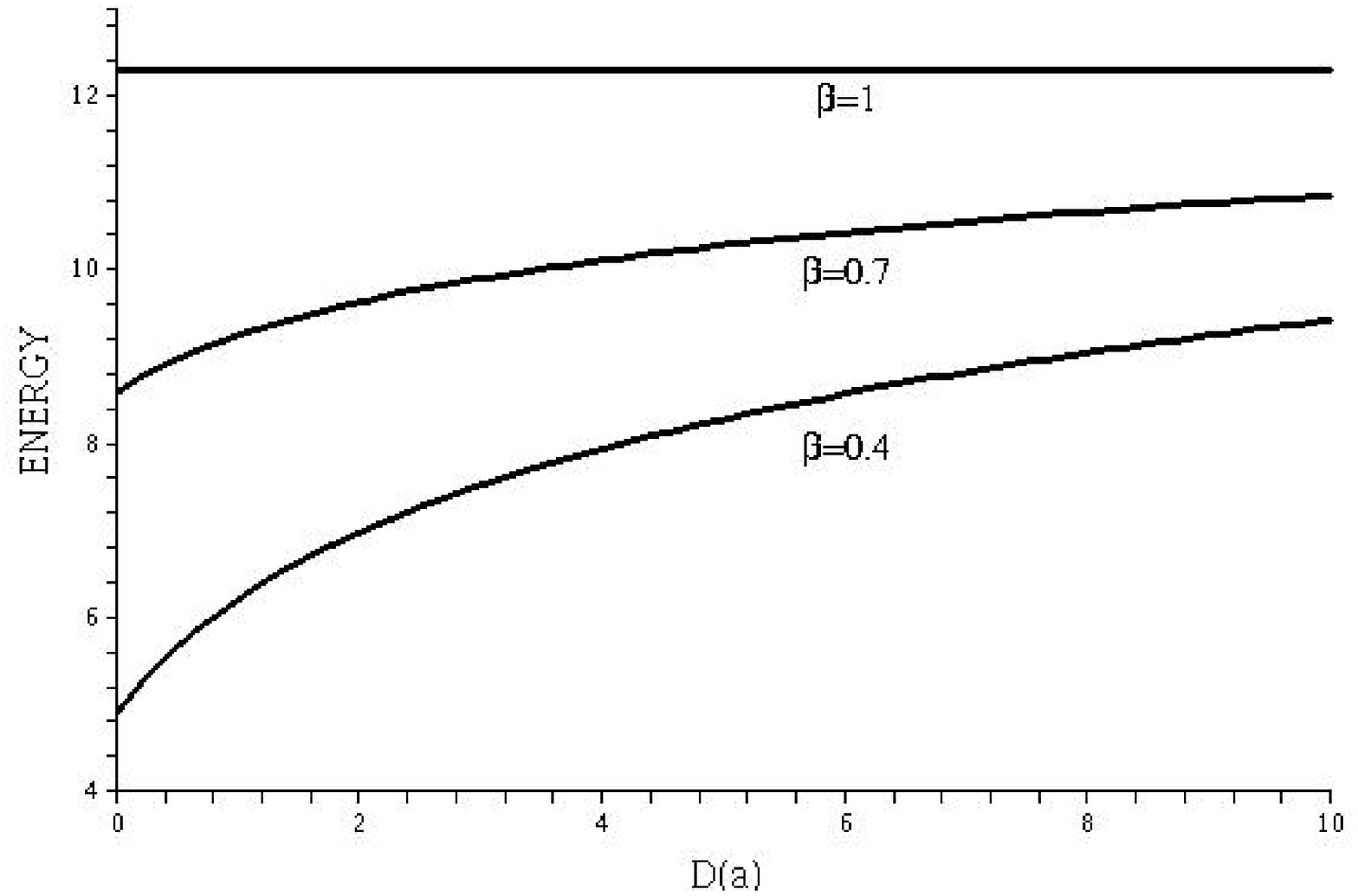}}
\vskip -0.0cm \caption{{\protect\small The energy of a single vortex on a cone as function of its center distance from the tip, $D(a)$. We have taken $J,Q$ and $a_0$ equal to the unity while $L=50a_0$.}}
\label{figure2}
\end{figure}

In the preceding section, our analysis was performed for a vortex
centered on the conical apex, where we obtained the energies of the two possible solutions. We may wonder whether these are the
unique possible cases. Actually, as we shall show below
these are the most energetically favorable configurations but not the
unique ones. As we have seen, outside the apex region spins tend to be encountered lying on the
surface. Therefore, we shall explicitly deal hereafter with the in-cone excitations. In order to
see this, let us consider
an in-cone magnetic vortex whose center
is apart $D$ from that point which, after identification of lines,
${\bf u}={\bf v}$, will play the role of the conical apex.
Furthermore, it should be emphasized that the cylindrical symmetry
of the vortex must be verified according to ${\bf n}$-vector. Such
a configuration centralized at the point $(x_{0},y_{0})$ on the
plane with a wedge, reads (up to a constant),
$\Phi_{D}=\arctan[(y-y_0)/(x-x_0)]$, whose energy may be
explicitly calculated to be:
\begin{equation}
E(D,\beta)=Q^2\int^{D+{a}_{0}} _{{a}_{0}}\int^{2\pi}_{0}\,
\frac{dr}{r}\,d\varphi + \int^L_{D+a_0}\int^{2\pi\beta}_{0}\,
\frac{dr}{r}\,d\varphi=\pi Q^2\,J\ln\left[
\frac{L^\beta\,(D+a_0)^{(1-\beta)}}{a_0}\right]\,. \label{EnergyD}
\end{equation}
where $\varphi$ measures the relative angle of the spins relative to an axis joining the 
point $\vec{P}$ and the origin. Essentially, as illustrated in Figure \ref{cone2}, all the spins inside a disc of radius $D$, centered at $P$, are kept after identification of lines, so that no angle deficit effect is experienced by this term. However, a number of the outer sites should be removed from energy considerations, what gives rise to the factor $2\pi\beta$ in the angular integration of the second term, above. [The vortex core radius, $a_0$, is introduced to avoid spurious divergence in the calculation].\\
Some limits are in order: first, note that as long as $\beta\to1$
then $E(D,\beta=1)=\pi\,J\ln({L/a_0})$ recovering its flat
counterpart. On the other hand, whenever $\beta\to0$ then $D$ is
meaningless and vanishes, so that $E(\beta=0)=0$. Now, if we set
$D=0$ we recover equation (\ref{Eic}), as expected. However,
as long as $D$ becomes very large, $D\to\,L\to\infty$, vortex
energy is unaffected by conical geometry, say,
$E(D\to\infty,\beta)=\pi\,J\ln({L/a_0})$. The last result may be
easily understood if we remember that the vortex energy density is
proportional to $r^{-2}$, so that, quite apart from the vortex core,
it is very small and so practically not sensitive to area deficit.
This is also the reason why its total energy is increased as long as
its center is moved away from the conical apex (Figure \ref{figure2}).

\begin{figure}[!h]
\centering \hskip 0cm
\fbox{\includegraphics[width=14cm,height=7cm]{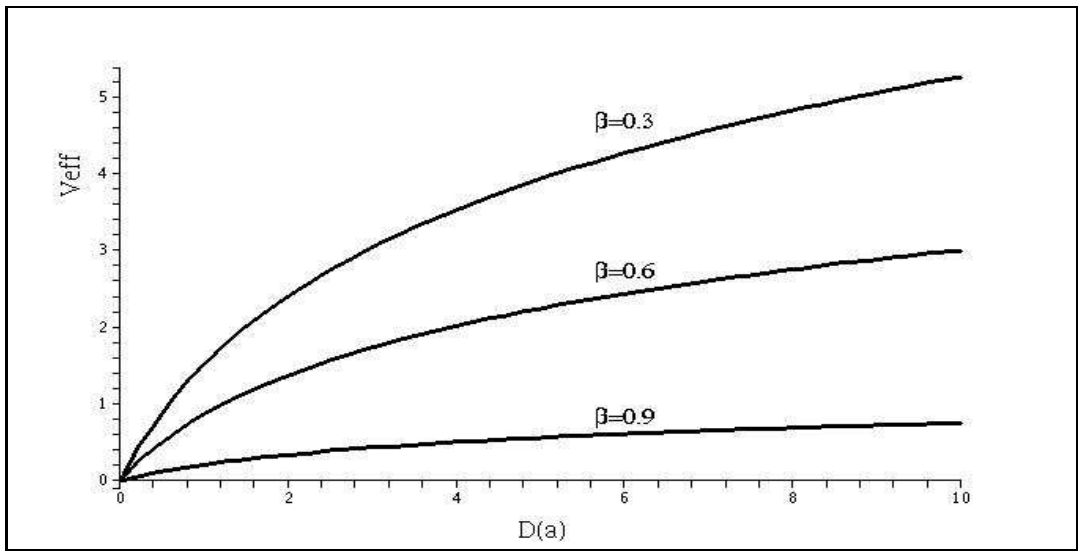}}
\vskip -0.0cm \caption{{\protect\small The effective potential of the apex on a in-cone vortex as function of its center distance from the tip, $D(a)$. We have taken $J,Q$ and $a_0$ equal to the unity while $L=50a_0$.}}
\label{figure3}
\end{figure}

Thus, we may understand that apex attracts magnetic
in-cone vortex according to
\begin{equation}
V_{\rm apex}(D,\beta)= E(D,\beta)-E_{\rm ic}=\pi
Q^2\,J(1-\beta)\ln\left[1+\frac{D}{a_0}\right]\label{Vapex}\,,
\end{equation}
so that in the limit $D/a_0 <<1$ we get a linearly confining
potential, $\pi\,J (1-\beta) D/a_0$. It should be emphasized that,
although $a_0$ is the smallest cutoff length scale of the model, the
limit $D/a_0 <<1$ simply states that whenever the vortex is pinned
to the apex cone, a strong attractive potential appears against
dislocations of its center from that point (Figure \ref{figure3}). This is to say that, if
the vortex center could develop small oscillations around the apex,
it would experience the linear confining potential
presented above. For the sake of comparison, a similar small
oscillation of a vortex around spin vacancies was predicted for
diluted easy-plane magnets \cite{PereiraJFK05}. Here, such a
scenario resembles quark confinement, where a linear potential
provided by gluons keeps quarks together. In addition, in the limit
$\beta\to0$ such a potential is not continuous since a naive
calculation gives $V_{\rm apex}(D,\beta=0)=\pi\,J\ln(1+D/a_0)\neq0$.
However, as stated above, $D=0$ as long as $\beta=0$, so that this
potential must vanish in this case. Therefore, expression
(\ref{Vapex}) is explicitly valid only for non-degenerate cone, say,
$0<\beta\le1$. Such a defect in the space may also cause important
effects in the spin dynamics and configurations. Really, it is
expected that the spins in the neighborhood of the cone apex, which
is a singularity, must have larger fluctuations than the distant
spins. Then, it is conceivable that an in-cone vortex reaching the tip should change to an
out-of-cone excitation, unless $\delta$ be strong enough to keep them on the cone surface. Furthermore, the translational symmetry is broken on this surface since
there is a preferential position for the nucleation of these excitations.  \\

Because apex nucleated magnetic vortices appear to be stable states, we may
wonder whether this could be thought as a mechanism for magnetic
recording and logic gates, for example, in the lines put forward
in Refs. \cite{Ross-nanocone,Rossetal,Rahmetal}. Furthermore,
following ideas of Refs.\cite{Rahmetal}, we may wonder whether
conical shaped defects could not be incorporated into submicron
magnetic particles so that vortices would appear nucleated around
conical apices instead of hole-pinned vortices metastable states.
Besides, there is a similarity between the system studied here and
magnetic force microscope tips, where it is well known that a
highly pointed tip implies in localization of a strong
magnetic field there. Also in this line, we could combine tips and nonmagnetic impurities, which has also been shown to attract topological objects\cite{vacancy-th,Molprb66,AfranioPiresJMMM,AfranioJFprb68,Wysin2005q2,AfranioJMMM279,Wysinetal2005,vacancy-ex,Afranio2005nano}. Then, a scenario with more than one kind of attracting point may give rise to externally controllable multistable states of vortices nucleated around tips or impurities.\\

\section{Two vortices on the cone}
At some extent, a vortex is similar to an elementary charge in
electrostatics, even in a conical surface. Indeed, the self-energy of both objects blows up at
both $\rho\rightarrow 0$ and $\rho \rightarrow
\infty$ limits, like in the usual flat plane system. Actually, the
singularity at $\rho =0$ may be regularized by an additional
short-range field ($m(\rho)$) to form a finite vortex core, but for a
strong enough anisotropy, the spins tend to lie on the surface even
in the region of the vortex core (around a plaquette). However, the
logarithmic divergence for large $\rho$ remains so that a single
vortex energy blows up in an infinite system (here, an infinite
cone). Therefore, other vortices must be present. For instance, if a
vortex appears together with an antivortex (to form the equivalent
of a dipole) then the pair energy is finite. Thus, a natural
question to be answered is how vortices interact on the conical
surface. For that, let us consider the case of two vortices, with
charges $Q_1$ and $Q_2$, on the cone. Here, we shall assume that $\delta>>1$, so that only in-cone vortices are expected to be found (see Section 2). [The scenario with transition to out-of-cone profiles very close to the apex could be considered, at principle, in the same lines]. We use $x-y$
coordinates of a plane with a deficit angle (we take the origin to
be at conical apex). Their profile reads like follows: \ba \Phi_{\rm
Q_{1}Q_{2}}=Q_1 \arctan\left(\frac{y-y_v}{x-x_v} \right)+Q_2
\arctan\left(\frac{y-y_{\tilde{v}}}{x-x_{\tilde{v}}}
\right)\,,\label{vortex-pair} \ea whose energy may be analytically
carried out and reads:
\ba
E_{Q_{1}Q_{2}}(D,\beta)=\pi^2\beta\,J(Q^2_1+ Q^2_2)/2+\pi\beta\,
J(Q_1+Q_2)^2\ln(L/a)+V{\rm eff}\,,\label{Epaircone}
\ea
Here, the first term is the energy formation of vortices while the second answers for their
energies according to the sample
size, $L$. In addition, the effective potential,
$V_{\rm eff}$, is given by:
\ba
V{\rm eff}(Q_i,D,\beta)=-2\pi\beta\,
J\,Q_1Q_2\ln({\cal D}/a)+\sum_i
\pi\,J(1-\beta)Q^2_i\ln\left(1+\frac{|\vec{D}_i|}{a_0}
\right)\label{Veff2qapex}\,,
\ea
where $\vec{D}_i$ ($i=1,2$)
localize the vortices separated by ${\cal D}$ (the less path joining
the two vortices, generally appearing to be an arc segment on the
cone). [Note that, as long as $\beta\to1$ usual plane-like results
are recovered]. Clearly, the first term in eq. (\ref{Veff2qapex}) appears to be repulsive
(attractive) if the vortices present charges of the same (opposite)
signals, like usually (but depending on
$\beta$-parameter). Thus, a vortex and an
antivortex attract each other while two vortices (or antivortices)
mutually repeal. The second term is traced back to the attractive
effect of the conical tip on a vortex (or antivortex). Therefore, we
have two cases depending upon the signals of the topological
charges.\\

\begin{figure}[!h]
\centering \hskip 0cm
\fbox{\includegraphics[width=14cm,height=7cm]{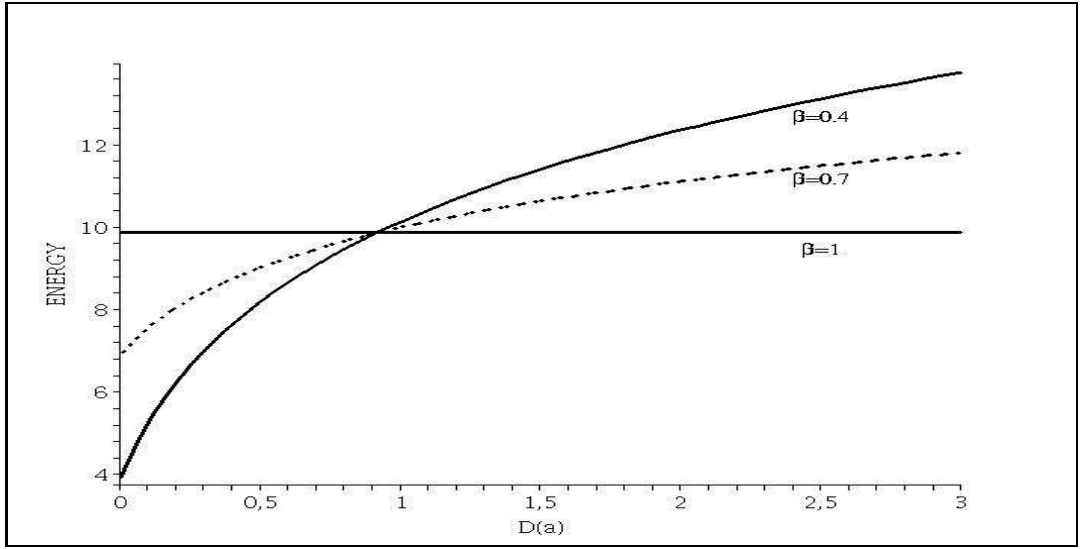}}
\vskip -0.0cm \caption{{\protect\small Shows how the vortex-antivortex pair energy behaves
as function of $D(a)$. Clearly, the pair is expected to nucleate around conical apex, $D=0$ [we
have taken ${\cal D}=a$, $J=1$].}}
\label{E-pair-conefig}
\end{figure}

First, if vortices have opposite charges, then they are
expected to be attracted to the apex. This fact is illustrated in
Figure \ref{E-pair-conefig} which displays how energy of a
vortex-antivortex pair with unity charges, $Q_1=-Q_2=+1$, behaves
as function of distance vortex-apex, $D(a)$, for some values of $\beta$ parameter (the antivortex
is ${\cal D}=a$ apart from the vortex, $Q_1$). Notice
that as long as cone is narrowed the pair
energy appears to get lower values. In addition, note also that
the potential well provided by the apex gets deeper, say, the stable
states for the pair correspond to one of the vortices pinned
to the tip. Analogously to the single vortex case, a geometrical
pinning of magnetic vortex is also observed
here.\\

On the other hand, whenever both vortices carry the same signal
charges, for simplicity, we take $Q_1=Q_2=+1$, then $V_{\rm eff}$,
(\ref{Veff2qapex}), consist of competing potentials: vortices
repealing each other while the apex working to keep them together,
near the tip. Actually, on the conical surface the apex attraction
can dominate mutual repulsion so that a system with two vortices
with the same charge may, at principle, nucleate around the tip. In
order to investigate such a possibility in more details, let us fix
some configurations for the vortices on the cone. First,
let us take both of the two vortices at
the same distance from the apex, $|\vec{D}_1|=|\vec{D}_2|\equiv D$,
but centered at diametrically opposite points (thus, separated by
${\cal D}=2D$ if $\beta> 2/\pi$ or ${\cal D}=\beta\pi D$ whenever
$\beta\le 2/\pi$). For this configuration, the minimum of $V_{\rm
eff}$, eq. (\ref{Veff2qapex}), will occur, for a given $\beta$, at:
\ba D_{\rm
min_{1}}=\frac{a_0\beta}{(1-2\beta)}\,,\label{Dmin1}
\ea
which states that a finite value take place only for $\beta<1/2$ (cone
aperture half-angle $\alpha<30^o$). In words, given a $\beta<1/2$
there exists a distance from the apex, $D_{\rm
min_{1}}$, which supports a stable state of the two vortices bounded to the apex.\\

\begin{figure}[!h]
\centering \hskip 0cm
\fbox{\includegraphics[width=14cm,height=7cm]{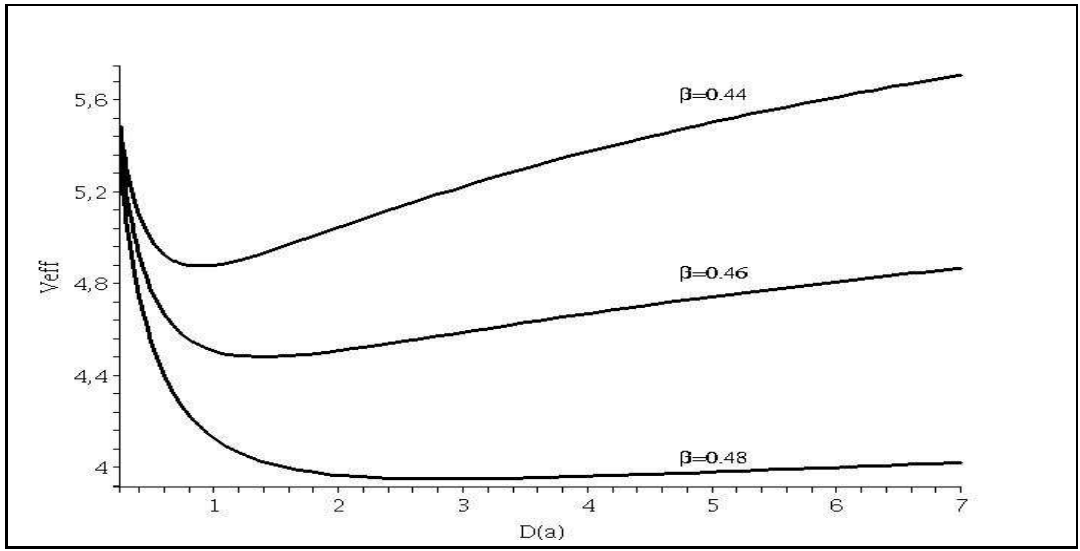}}
\vskip -0.0cm \caption{{\protect\small Shows how $V_{\rm eff}/J$ behaves
as function of $D(a)$ for the case $Q_1=Q_2=+1$. Note that, in despite of the vortices have the
same charge, energetics tends to keep them tied near the conical tip. [The values for $D_{\rm
min_{1}}$ read around $0.88a$, $1.38a$ and $2.88a$ for $\beta$ equals to $0.44$, $0.46$ and $0.48$,
respectively].}}
\label{2vortices}
\end{figure}

Indeed, for $\beta\lessapprox 1/(2+a_0)\approx0.447\equiv{\beta}_{\rm cr_{1}}$ ($\alpha\lessapprox
26.55^o$)
stability demands $D<a$. [From the result above for ${\beta}_{\rm cr_{1}}$ we clearly realize
that vortices-apex nucleation possibility is also intimately related to the finite size of the
vortex core, $a_0$]. In this case, we would have effectively both vortices at the tip (in the
discrete scenario). Energetics could, in this case, favors the formation of an unique $Q=2$
excitation instead of two $Q=1$ vortices. In this line, we should mention the work of
Ref.\cite{Wysin2005q2}, in which doubled charged vortices appear nucleated around spin vacancies
in the standard plane,
as demonstrated by simulations, by virtue of the attractive effect of such impurities. [Although
our
results cannot be extrapolated to short range, say $D\lesssim{a}$, it would be interesting to
simulate a similar system on the conical surface]. On the other hand, whenever $\beta\gtrapprox
0.447$, the two vortices nucleate around the tip with $D\geqq a$. This could be viewed as a
vortices-tip bound state, since the effective potential presents a global minimum at $D_{\rm
min_{1}}$. Let us recall that such a possibility is ruled out from usual planar-like systems.
Figure
\ref{2vortices} displays how the effective potential behaves as function of $D(a)$ for some
values of $\beta$. As may be clearly realized, as long as
$\beta$ increases from ${\beta}_{\rm cr_{1}}$ to $1/2$ then $D_{\rm min_{1}}$ shifts to higher
values.

Another possible configuration is that in which one vortex is already centered at the apex while
the another is apart $D$. In this case, one of the terms presented in the sum of eq.
(\ref{Veff2qapex}) identically vanishes. The remaining potential is minimized at:
\ba
D_{\rm min_{2}}=\frac{2a_0\beta}{(1-3\beta)}\,.\label{Dmin2}
\ea
Now, stability is possible only for narrower cones, say $\beta<1/3$ ($\alpha\lessapprox 19.47^o$).
However, according to the relation above, vortices-apex stable states with $D\ge{a}$ demands
$\beta\lessapprox 0.287\equiv\beta_{\rm cr_{2}}$. In summary, we conclude that the nucleation of
two vortices (or antivortices, as well) around the conical tip is possible for some narrow
apertures. If the two objects are separated from the apex of $D=D_{\rm min_{1}}\ge{a}$, then cone
aperture half-angle $\alpha\equiv\alpha_1\in(26.55^o,30^o)$. However, if one object is already
nucleated at the tip and the another is $D=D_{\rm min_{2}}\ge{a}$ apart, a narrower cone is
demanded, say $\alpha_2\in(16.68^o,19.47^o)$, approximately.\\

\indent Let us now consider the problem of vortex-pair dissociation in this background. Our estimative will consider only in-cone vortices. First of all, let us emphasize that it is nowadays well-established that the critical temperature, of vortex-antivortex depairing, is $\approx
0.284 \pi J/k_B$\cite{Olsson9495}, within the planar rotator, while for the XY-model, where spins can have out-of-plane fluctuations, it reads $\approx 0.22\pi J/k_B$
(see Ref. \cite{XYTBKT} for
details). However, it has been recently raised the question of how vacancies affect this transition (see, for example,
\cite{AfranioJFprb68,Wysin2005q2,Wysinetal2005,Leonel03} and related
references therein). Actually, several simulations have shown that
vacancies tend to diminish the transition temperature, so that at
vacancy concentrations $\varrho$ equal or above the percolation
threshold $\varrho_{pt} \approx 0.41$, the topological transition
is ruled out at any finite temperature\cite{Wysinetal2005}.\\

As we have noted in the introduction, the deficit
angle can be viewed as a region with nonmagnetic impurities
grouped in a cluster with a wedge shape (see Figure \ref{cone}).
In this case, the impurity concentration would be proportional to
$2\pi\gamma$, i.e., proportional to the deficit angle (the problem
with this interpretation is the boundary conditions required at
the edges of the wedge). To
estimate the critical temperature for the conical surface we use
the same argument of energy-entropy as Kosterlitz-Thouless performed for
the standard plane\cite{KT}. Here, the system area is given by $\beta L^2$
and then, the entropy may be estimated to be\footnote{Since the apex is more energetically
favorable than other points, we have a nonuniform probability distribution over the surface.
Thus, our estimative is good if a small area around the tip is not considered. Therefore, our
analysis is strictly valid only for in-cone excitations, whose centers are expected to be moved
away from the apex.} $S_{\rm cone}=k_B
\ln(\beta{L}^2/\beta{a}^2)=2k_B\ln(L/a)$, like in the usual planar
case. Therefore, a rough analysis gives the following range for
depairing temperature of vortex-antivortex excitations in the $XY$
model defined on a conical geometry
\begin{equation}
\pi{\beta}J\,<k_B\,T_{\rm cr-cone}< \pi
J\,\label{TBKT-cone}.
\end{equation}
Actually, if we could extrapolate results of Ref. \cite{XYTBKT} to
the present geometry, we would get
$0.22\,\pi{\beta}J < k_B\,T_{\rm cr-cone} < 0.22\,\pi J$. Therefore, the critical temperature $T_{c}$ of the system have a linear
dependence on the geometrical parameter $\beta$.

\section{Conclusions and Prospects}
In the present work we have considered vortex-like excitations of
$XY$-like model on a circular conical support. Such a space is locally
Euclidian, since a conical singularity is created after removing a
wedge of the usual plane and gluing its edges. Vortex solutions and
configurations depend on vortex position and deficit angle in a
nontrivial way so that our analysis of this problem has not been
rigorous. Nevertheless, the results presented here may be of
interest for many condensed matter systems (2D electrons gas on the cone) and field theories on
manifolds with nontrivial geometry as well as for lower dimensional
gravity, in which the anisotropic nonlinear $\sigma$ model
and its associated vortex-like solutions are coupled to the Einstein field. We have seen
that whenever lying on a cone, vortex (or antivortex) energy appears
to linearly decrease with conical aperture angle, so that as cone is
extremely tightened ($\beta\to0$), its energy vanishes like sample
size does. We have studied in-cone and out-of-cone vortex excitations. For suitable ranges of the
anisotropy, the first type appear to be found far away from the tip while the second ones tend to
be centered at the apex. In addition, we have seen that it is energetically
favorable for a magnetic vortex to nucleate around the conical
apex. Thus, an in-cone could become an out-of-cone vortex as long as it is trapped by the tip.
Therefore, we could think conic-like magnets as pinning
defects for such excitations.\\

Going further, we have taken two vortices on the cone. Analogously
to the former case, even though its energy has a logarithmic
dependency on their distances, it also depends linearly on $\beta$
parameter. In the case of a vortex-antivortex pair
($Q_1=-Q_2=+1$), our results show that it is energetically
favorable for nucleation around the apex, like in the single
vortex case. An
intriguing result emerges as long as we take $Q_1=Q_2$.
Although repealing each to other, like in the usual planar-like
case, conical apex tends to attract and keep them around the tip.
Actually, we have realized that for conical aperture smaller than
a critical value, apex potential dominates repulsion, which leads
to the nucleation of the vortices. As a final result, connecting geometrical aspects with
Berezinskii-Kosterlitz-Thouless transition related issues, on a
cone, such a transition is expected to takes place at lower
temperatures.

As Prospects for future investigation we may quote, for example:
i) simulations of spin models in order to confirm the assertions
and results presented here; ii) study the ``out-of-cone vortices"
and possible other deformed solution such as spiral vortices
\cite{Borisov05}, which could be present in non-circular cones;
iii) consider the spin dynamic in curved spaces and/or in surfaces
containing defects\cite{PereiraJFK05}; iv) study of
magnetostatic energy on the cone surface, mainly in connection with nanosized magnets. \\

Finally, our present study could be also useful for some related systems, like superconducting vortices, dislocations, and so forth, whenever they lie in conic-like geometries.
\vskip 1cm
\centerline{\large\bf Acknowledgements} \vskip .3cm
The authors thank CAPES, CNPq and FAPEMIG for financial support. WAM-M
also thanks DCB/UFVJM where part of this work was done.\\ 

\end{document}